\documentclass[%
 reprint,
 showkeys,
 amsmath,amssymb,
 aps,
]{revtex4-1}

\usepackage{graphicx}
\usepackage{dcolumn}
\usepackage{bm}

\begin{document}

\preprint{APS/123-QED}

\title{Interference with non-interacting photons and a special type of detector}

\author{Ioannis Contopoulos}
 \email{icontop@academyofathens.gr}
 \affiliation{Research Center for Astronomy and Applied Mathematics, Academy of Athens, Athens 11527, Greece}




\date{\today}

\begin{abstract}
We present a statistical model of non-interacting individual classical particles that may lead to a microscopic implementation of quantum mechanics. The model requires the action of a special type of detector that detects and records individual particles. According to our model, microscopic particles are classical entities that follow classical trajectories without nonlocal interactions between them. As long as they remain undetected, particles carry the information of a phase equal to an action integral along their trajectory. At the point of their detection, our special type of detector collects the phases from all particles reaching it, adds them up over time as complex numbers, and divides them by the square root of their number. The detector records a detection each time the square of the magnitude of the ensuing vector crosses an integer value. We test our model with numerical simulations of a double-slit experiment, and obtain an interference pattern analogous to the one observed in an actual physical experiment.

{\bf Statements and Declarations:} The datasets generated during during the current study are available from the corresponding author on reasonable request. There are no ethical issues nor external funding associated with this work. On behalf of all authors, the corresponding author states that there is no conflict of interest. 
\end{abstract}

\maketitle


\section{\label{sec:level1}Introduction}

Quantum mechanics presents a conceptual conundrum that led to a multitude of interpretations, all of which are, in our opinion, unsatisfactory \cite{1,2,3,7}. Individual measurements detect individual particles, but the event-by-event collection of such measurements shows wave-like interference patterns which, quoting Feynman, are `impossible, absolutely impossible to explain in any classical way and have in them the heart of quantum mechanics' (Feynman Lectures on Physics \cite{4}). We will argue, as did Jin et al.~(2010 \cite{5}) a few years ago, that it is wrong to think that waves are the only possible physical cause of interference. 

We will first show that it is possible to replace the collective description of a wave with a sequence of non-interacting individual particles if each particle carries a phase that evolves along the particle trajectory. Not only that. In order to observe wave effects, we need the contribution of a particular type of classical detector, namely one that records the phases of the detected individual particles. In \S~II we derive the operating characteristics of such a detector in the case of photons, without any reference to Planck's constant. In \S~III we show how we can obtain an interference pattern from the collection of a sequence of such non-interacting individual photons. This will further elucidate the role of our detector. In \S~IV we try to generalize our conclusions for massive particles with mass $m$, velocity ${\rm v}$. We will argue that the classical description of the trajectories of microscopic particles with energy $\epsilon$, and potential energy $U$
is missing something fundamental about their motion. This may be that particles undergo stochastic interactions with a background, or with the detectors that result in their measurement. 
We summarize our conclusions in \S~V. The novel result of our work is that, if physical measurement operates according to the characteristics of the special type of detector described in \S~II, 
the interference pattern is a-posteriori collective information that is obtained from the collection of a growing number of particle detections, and is not known by each individual particle.

\section{\label{sec:level1}Collection of the phases of non-interacting photons}

Our eventual goal is to obtain a physical interpretation of quantum mechanics based on individual non-interacting particles that behave classically. Although we will eventually consider classical particles like electrons, it will help our discussion to begin by considering individual photons. Let us discuss first what information individual photons may carry that can lead to the construction of event-by-event interference patterns at a detector. 

According to classical electromagnetism, an electromagnetic wave in vacuum is characterized by its electric field, its direction, and its frequency $\nu\equiv c/\lambda$ ($c$ is the speed of light, and $\lambda$ is the wavelength of the radiation). These are the most fundamental quantities from which one can calculate the Poynting flux of energy and the electromagnetic energy density. Quantum mechanics offers a different picture of the electromagnetic wave in terms of individual photons. If we were to reconcile the wave and particle pictures without invoquing non-local interactions between photons, the flux of energy may be viewed as the flow of $N$ photons per unit time each carrying energy $\epsilon$ at the speed of light along the direction of the Poynting vector. We will now argue that non-interacting photons cannot carry the information of the electric field which is fundamental for the manifestation of wave behavior (e.g. refraction, diffraction, interference, etc.). The reason is that the collection of $N$ indistinguishable photons carries $N$ times the energy $\epsilon$ of one photon, but the amplitude of the electric field of the corresponding classical wave is proportional to only $\sqrt{N}$ times the square root of $\epsilon$. Taking the square root of $N$ is obviously a collective operation that cannot be carried as information by non-interacting individual particles.

We will now propose one particular way to count the number of photons in a detector. Firstly, we will assume that individual photons carry an evolving phase $\phi$, something like an `internal clock'. 
For photons of wavelength $\lambda$ we propose the natural expression
\begin{eqnarray}
\phi &\equiv & 2\pi\frac{{\rm l}}{\lambda}
\ .
\label{phiphoton}
\end{eqnarray} 
Here, 
${\rm l}$ is the length of the photons' trajectory from their point of origin to the finite region of spacetime where and when their detection takes place. Secondly, we will assume a special type of detector that collects the phases of the $N$ photons that reach it at a finite region around the spacetime position $({\bf x};t)$, and forms the expression
\begin{equation}
\overline{\Psi}({\bf x};t) \equiv \sum\nolimits_{k=1}^K\frac{1}{\sqrt{N_k}}\sum\nolimits_{j=1}^{N_k} {\rm e}^{2\pi {\rm l}_k/\lambda}
\label{Psi}
\end{equation}
Hereafter, $\overline{(\ldots)}$ will denote quantities that are calculated after several repetions of an experiment. $k=1$ to $K$ denotes the $K$ possible different trajectories that photons may follow to reach the spacetime position $({\bf x};t)$, as e.g. in an interference experiment, and $l_k$ are the lengths of every such trajectory. $N_1+\ldots + N_K=N$ is the total number of photons collected by the detector at spacetime position $({\bf x};t)$. We will discuss the physical significance of eq.~(\ref{Psi}) below. Finally, the number of photons {\it recorded} by the detector in that particular region of spacetime around $({\bf x};t)$ is assumed to be equal to
\begin{equation}
\overline{N}\equiv |\overline{\Psi}|^2({\bf x};t)=\left|\sum\nolimits_{k=1}^K \frac{1}{\sqrt{N_k}}\sum\nolimits_{j=1}^{N_k} {\rm e}^{2\pi i{\rm l}_k/\lambda}\right|^2\ .
\label{N}
\end{equation}
In general, the number $\overline{N}$ of recorded photons is {\it not equal} to the number $N$ of collected ones. This is the main ansatz that characterizes our process of photon collection and recording:
\begin{quotation}
\noindent
{\it We assume that the only way to count photons is via eq.~(\ref{N})}.
\end{quotation}
In other words, $N$ photons are collected at a particular position in spacetime, but the detector says that $\overline{N}\neq N$ have been recorded. The important thing to emphasize here is that $\overline{\Psi}$ is not known a-priori, and is certainly not `known' by each individual photon. $\overline{\Psi}$ is determined at the detector after the collection of a large number of photons from subsequent repetitions of the experiment. 

We will now see how eq.~(\ref{Psi}) allows us to generate an interference pattern with individually collected non-interacting photons. 

\section{\label{sec:level1}Event-by-event emergence of photon interference}

Let us consider a double-slit experiment where individual photons of energy $\epsilon$ emerge horizontally from a certain source, cross a screen with two parallel perpendicular slits separated by a distance $D$, and then hit a horizontal detector some distance $l\gg D$ behind the screen. The detector consists of an array of pixels where individual photons are collected and their phases are processed according to eq.~(\ref{Psi}). The detector then triggers detections (i.e. records photons) every time $|\overline{\Psi}|^2$ crosses an integer value. The process repeats itself over and over again.

\begin{figure*}
 \centering
 \includegraphics[width=0.8\textwidth]{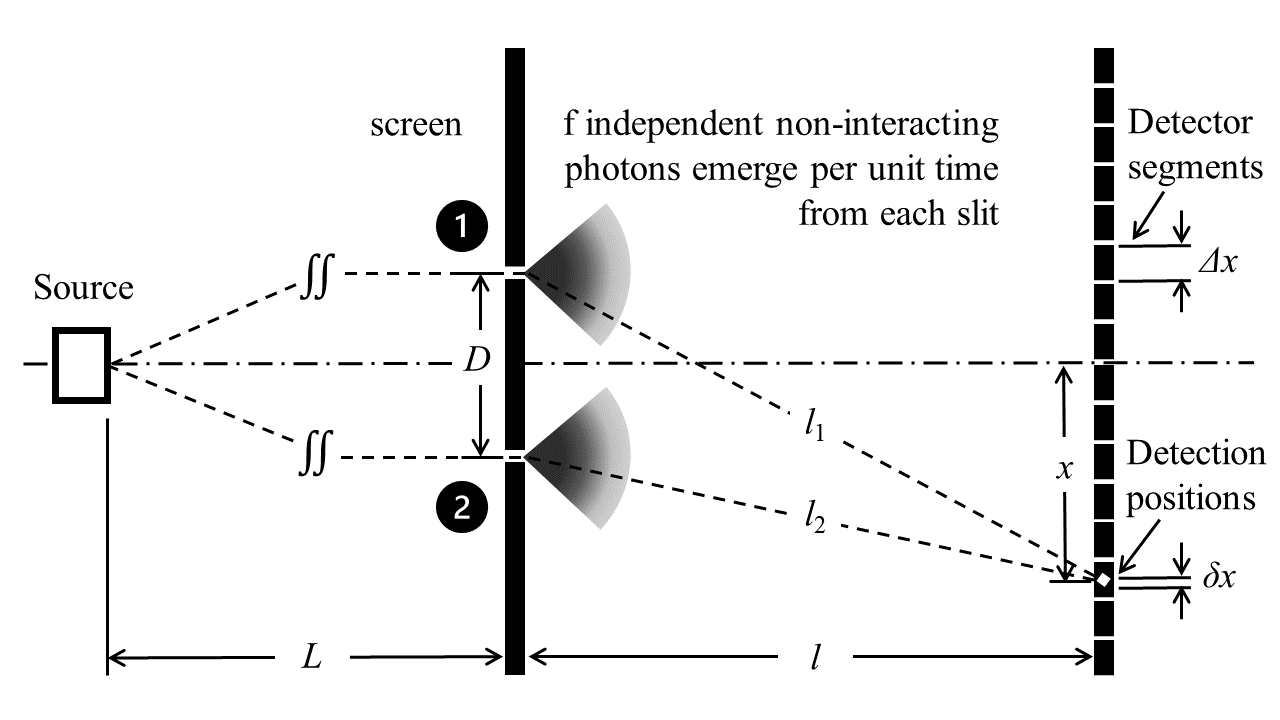}
\caption{Schematic of a double slit experiment with $f$ individual non-interacting photons emitted horizontally per unit time from each slit. $l,L\gg D$. Notice that there is no requirement for a common `coherent' origin of these photons. The only requirement is that their phases start counting from the time they emerge from the source.}
\label{figure1}
\end{figure*}

\begin{figure*}
 \centering
 \includegraphics[width=0.8\textwidth]{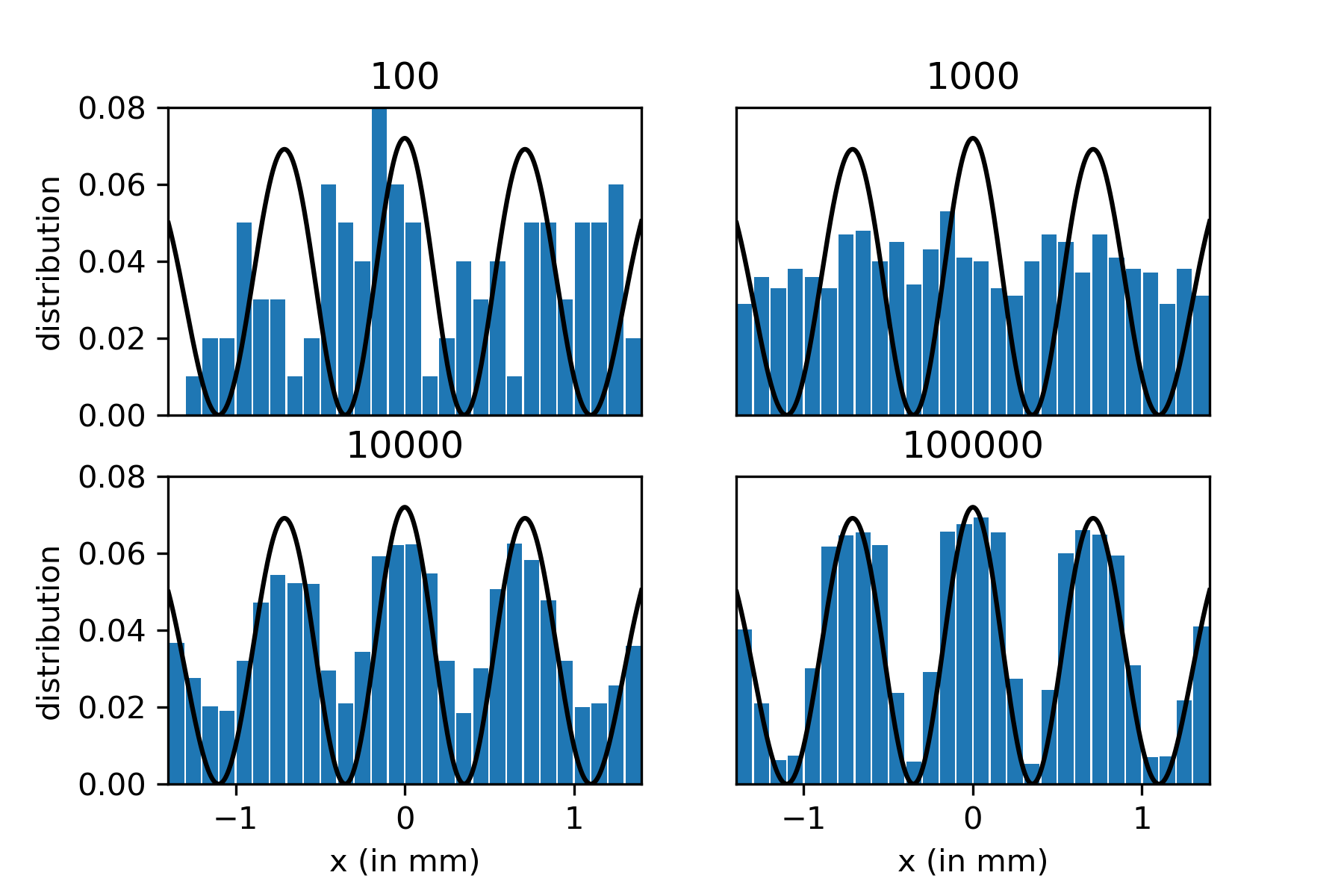}
\caption{Gradual buildup of the double-slit interference pattern for 100, 1,000, 10,000 and 100,000 photon detections respectively. Shown are the distributions of photons along the 28 $100\mu$m wide segments of the detector. In this experiment we chose detection pixels of width $\delta x=1\mu{\rm m}\sim \lambda$. Solid line: the analytical functional form of eq.~(\ref{Intphotons}). Note that the peaks drop very slowly with distance $x$ because photons are assumed to move on the horizontal plane, and therefore, their number density drops inversely to the distance from the slits, not the distance squared. $R^2>0.9$ after about 4,000 photon detections.}
\label{figure2}
\end{figure*}

\begin{figure*}
 \centering
 \includegraphics[width=0.8\textwidth]{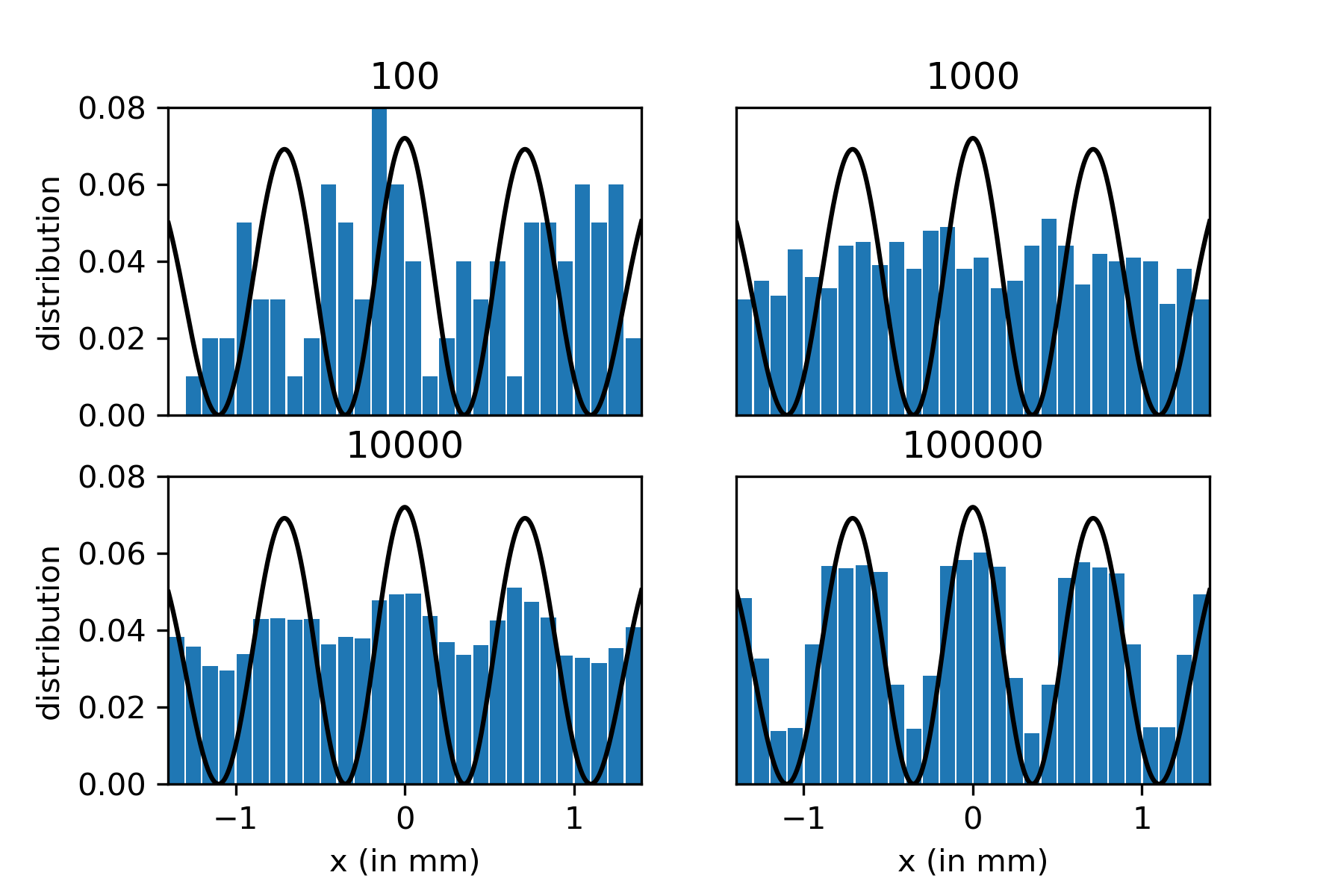}
\caption{Similar to figure~2 but for detection pixels of width $\delta x=0.2\mu{\rm m}< \lambda$. The interference pattern also develops clearly but more slowly than in figure~2.}
\label{figure3}
\end{figure*}

\begin{figure*}
 \centering
 \includegraphics[width=0.8\textwidth]{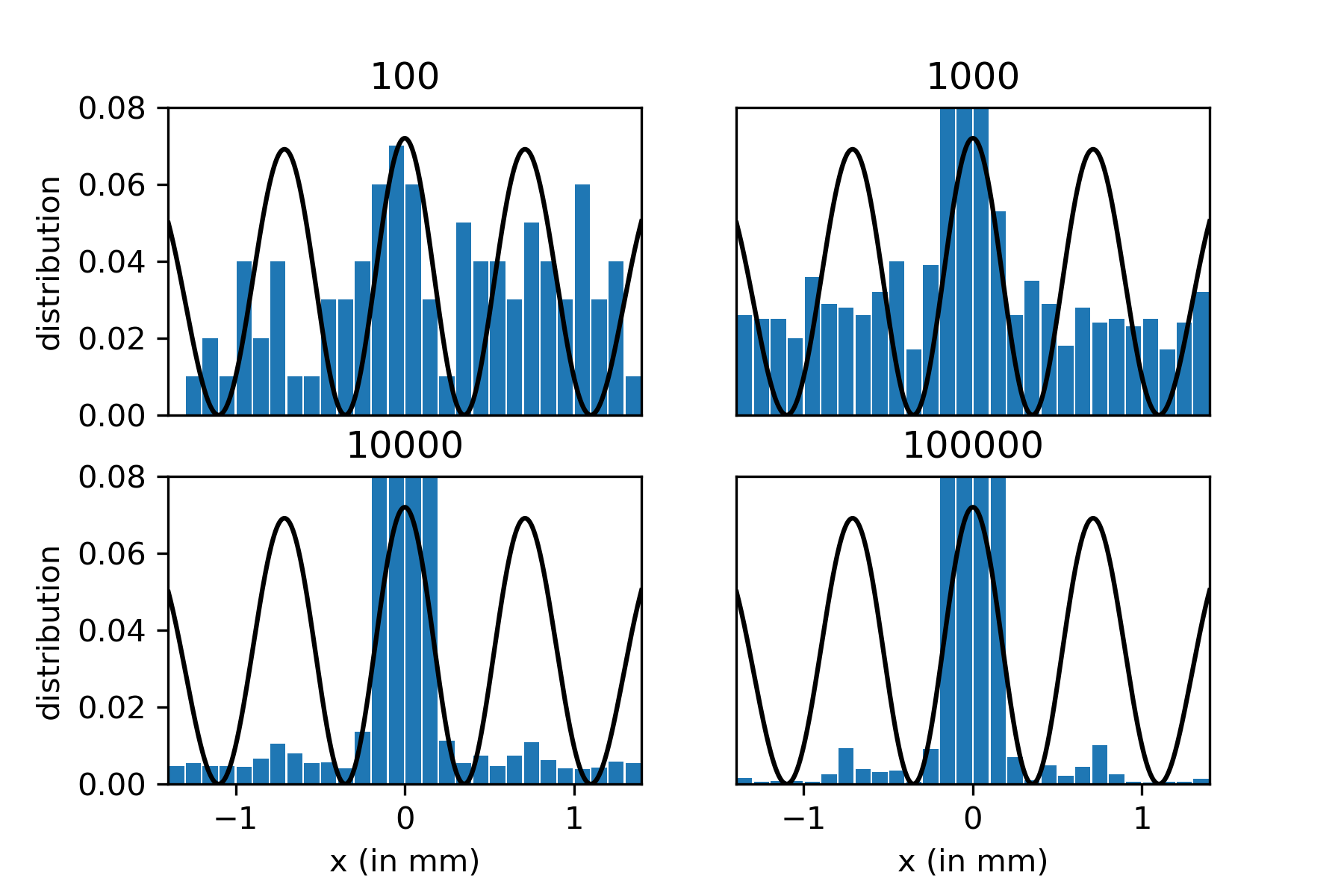}
\caption{Similar to figure~2 but for detection pixels of width $\delta x=10\mu{\rm m}\gg \lambda$. As expected, however long we may wait, the interference pattern will never be recovered.}
\label{figure4}
\end{figure*}

Let us assume that photons emerge at a rate of $f$ photons per unit time from each slit at the speed of light distributed isotropically on the horizontal plane, and that each detection position has a width $\delta x$ at positions $x$ from the detector midpoint (see figure~1). In reality, the detection position is a finite region of spacetime and not just one point. In other words, in order for our detector to process phases according to eq.~(\ref{Psi}), we need to consider a finite photon collecting area, and not just one point. As we will see in the following subsection, in order to obtain an interference pattern from the collection of a reasonable number of photons, the detection pixel width $\delta x$ must be neither too small, neither too large.

We also assume that there are no other external perturbations acting on the photons, thus all photons move along straight lines during their flight from the source to their eventual detection at the detector. Each pixel receives on average $fl\delta x/\pi \cdot[1/((x-\frac{D}{2})^2+l^2)+1/((x+\frac{D}{2})^2+l^2)]$ photons per unit time, which for $D\ll l$ becomes approximately equal to $2fl\delta x/(\pi (x^2+l^2))$, i.e. each pixel receives about $N=fl\delta x/(\pi (x^2+l^2))$ photons per unit time from each slit. According to eq.~(\ref{Psi}), $K=2$ (i.e. there are 2 independent paths to reach each pixel), and the function $\overline{\Psi}$ recorded by the pixel at position $x$ over an integration time $t_{\rm int}$ is equal to
\begin{eqnarray}
\overline{\Psi}(x;t_{\rm int}) & \equiv & 
\frac{1}{\sqrt{N_1}}\sum\nolimits_{j=1}^{N_1}{\rm e}^{2\pi i(L+l_{1j})/\lambda}\nonumber\\
& & +\frac{1}{\sqrt{N_2}}\sum\nolimits_{\mu=1}^{N_2}{\rm e}^{2\pi i(L+l_{2\mu})/\lambda}
\ .
\label{Ecalc}
\end{eqnarray}
Here, $N_1$ and $N_2$ are the numbers of photons that passed through slits No 1 and 2 respectively (they are both approximately equal to $N= fl\delta x\ t_{\rm int}/(\pi (x^2+l^2))$), and $L+l_{1j}, L+l_{2\mu}$ are the respective lengths of the two trajectories from the source to the detector pixel. From this point on we can only proceed with a Monte Carlo numerical simulation (see subsection~III.A). We can proceed analytically only if we assume that we have roughly an equal number $N$ of $j$ and $\mu$ summations. 
In that case, eq.~(\ref{Ecalc}) yields
\begin{eqnarray}
|\overline{\Psi}|^2(x;t_{\rm int})
& \approx & \frac{1}{N}\left|{\rm e}^{2\pi i(L+l_1)/\lambda}\sum\nolimits_{j=1}^{N}\left(1+{\rm e}^{2\pi i(l_2 - l_1)/\lambda}\right)\right|^2\nonumber\\
&&\nonumber\\
& = & \frac{1}{N}\left(N\left|1+{\rm e}^{2\pi i(l_2 - l_1)/\lambda}\right|\right)^2\nonumber\\
&&\nonumber\\
& = & 2N [1+\cos(2\pi(l_1 - l_2)/\lambda )]\nonumber \\
&= & 4N \cos^2\left(\frac{\pi xD}{2\lambda \sqrt{x^2+l^2}}
\right)\nonumber\\
&\approx & \frac{4fl\delta x\ t_{\rm int}}{\pi (x^2+l^2)}\ \cos^2\left(\frac{\pi xD}{2\lambda \sqrt{x^2+l^2}}
\right)
\label{Ecalc2}
\end{eqnarray}
when $D\ll l$. Notice that our process of photon detection differs from that of Jin et al.~(2010)\cite{Jin}.
In the next subsection we will perform a numerical experiment to test our result. 

We must emphasize a very important point in eq.~(\ref{Ecalc}). The summations over $N_1$ and $N_2$ are performed over photons that reached the detector after crossing slit No 1 and 2 respectively. However, 
\begin{quotation}
\noindent
{\it the detector `knows' the direction each photon came from}
\end{quotation}
and processes separately photons reaching it having crossed slit No1 from those reaching it having crossed slit No 2. Otherwise, if it added the exponentials with the phases of all the photons that reach it and then divided by the square root of their total number, namely $\sqrt{2N}$, the normalization in front of eq.~(\ref{Ecalc2}) would have been one half of the correct value.

Eq.~(\ref{Ecalc2}) yields the following variation of intensity with distance $x$ from the detector midpoint
\begin{eqnarray}
\frac{{\rm d}I(x)}{{\rm d}x}&=& \frac{\epsilon |\overline{\Psi}|^2(x;t_{\rm int})}{\delta x\ t_{\rm int}}\nonumber\\
&=& \frac{4fl\epsilon}{\pi (x^2+l^2)}\cos^2\left(\frac{\pi xD}{\lambda \sqrt{x^2+l^2}}
\right)
\label{Intphotons}
\end{eqnarray}
when $D\ll l$. Notice that if we integrate eq.~(\ref{Intphotons}) over all $x$'s we obtain the total energy per unit time crossing the screen with the two slits, namely
\begin{equation}
I=\int_{-\infty}^{+\infty}\frac{{\rm d}I(x)}{{\rm d}x}\ {\rm d}x=2f\epsilon\ .
\end{equation}
We thus confirm that, despite the collection and processing of phases and the emergence of an interference pattern, the total number of collected photons is equal to the total number of photons that passed through the two slits. Had we performed the summation in eq.~(\ref{Ecalc}) differently (first adding all the exponentials, and then dividing by $\sqrt{2N}$) we would be missing one half of the photons.

The derivation in eq.~(\ref{Ecalc2}) is correct only in the limit of an infinite number of detection positions $x$, and an infinite number of collected photons. In reality, each detection position has a finite size $\delta x$, thus the collected phases 
contain some variability on the order of 
\begin{equation}
\delta \phi \sim 2\pi \frac{\delta x}{\lambda}\ .
\end{equation}
The larger $\delta x$, the more the phases are mixed, resulting in a worse interference pattern.  The smaller $\delta x$, the less the phases are mixed, resulting in a clearer interference pattern. At the same time, however, the detector needs to collect a larger and larger number of photons to populate all detection sites, thus the interference pattern takes longer to appear. As we will see the optimal conditions for the generation of an interference pattern that closely mimicks an actual physical experiment with the minimum number of photon detections require $\delta x\sim \lambda$. In other words, the detection size must be `tuned' to the wavelength of the photon that it is going to measure!

\subsection{\label{sec:level2}Numerical interference experiment analogous to a physical experiment}

Let us now perform a numerical simulation of an idealized photon double-slit experiment similar to the one performed by Kolenderski et al.~(2014 \cite{8}). In that experiment, they recorded individual photons as they were gradually building the interference pattern one by one. They also compared the  outcome of their quantum mechanical experiment with the model of Jin et al.~(2010 \cite{5}) and found that, although it too was able to reproduce the interference pattern, it did so about 10 times slower than the physical experiment.

We too consider a source of coherent light 
with wavelength
$\lambda=842\ {\rm nm}$.
The detector consists of 28 segments of width $\Delta x=100\ \mu{\rm m}$ aligned around $x=0$ at a distance $l=3\ {\rm mm}$ away from a screen with 2 narrow slits a distance $D=5\lambda$ apart ($\lambda,D\ll l$). This particular distance between the screen and the detector was chosen so as to observe 4 interference peaks within the 28 segments of the detector, as in the actual experiment. A schematic representation of the setup between the screen and the detector is shown in figure~1. 
We are also given the information that the pixels have a 5\% detection efficiency in that particular wavelength range. Photons are collected at each of the 28 detector segments. In reality, however, photons are collected at detection positions (pixels) of width $\delta x\ll \Delta x$ within each segment. As we will see below, we were able to reproduce the results of the actual physical experiment for $\delta x\sim\lambda$. If $\delta x\gg \lambda$, the phases of individual photons that arrive at each segment are mixed so dramatically that no interference pattern is ever obtained. 

We perform the numerical experiment as follows. We send photons that move randomly through either one of the two slits, emerge at random angles behind the screen, and are collected in one of the $28\times (\Delta x/\delta x)\gg 28$ detection areas (pixels) of width $\delta x$. The phases of the collected particles are processed as in eq.~(\ref{Psi}) at each detection position. When $|\Psi|^2$ becomes equal to or larger than 1 at a certain position, the first particle is detected at that position. The next threshold is $2$ which when crossed, a second particle is detected, and so on and so forth. We stop the experiment when a total number of 100, 1,000, 10,000 and 100,000 detections are recorded. We then distribute these detections among the 28 detector segments because this is the only information available about how particle detections are recorded in the actual Kolenderski et al.~(2014) experiment with which we want to compare our results.

In figures~2-4 we depict the statistics of our numerical experiment for $\delta x = 1, 0.2$ and $10\mu{\rm m}$ respectively. These show how the interference pattern builds-up for particular values of the detection position width. We clearly see that when $\delta x\gg \lambda$, the interference pattern is destroyed. We found that, when $\delta x\sim \lambda$, the interference pattern is reproduced with a coefficient of determination $R^2\approx 0.9$ after about 4,000 detector triggerings. This is 20 times higher than the corresponding number of detector triggerings in the actual experiment of Kolenderski et al.~(2014). The discrepancy, however, may not be as high as it seems. In Kolenderski et al.~(2014) it is stated that, out of the $2\times 10^6$ photons per second generated by their source, only 72,000 photons per second reach their detector, whereas detector triggerings take place at a rate of only $200/(115\ \mbox{ms})=1,700$ triggerings per second. This would imply that each of the 28 pixels of their detector collects on average roughly $72,000/1,700/28=1.5$ photons for every triggering it makes. This, however, is not consistent with the claimed pixel photon detection efficiency of 5\%. This information leads us to believe that their detector collected 20 times more photons, i.e. their experiment reached a coefficient of determination $R^2\approx 0.9$ after collecting about 4,000 particles. This value is consistent with our result, and corresponds to their stated 200 detector triggerings. Notice, however, that in order to record 4,000 triggerings in our numerical experiment with $\delta x=1\mu$m, our detector collected 3 times more photons. Bearing further testing and comparison with actual experiments, we would like now to discuss how the above may be generalized for massive particles.

\section{\label{sec:level1}Towards the wavefunction of massive particles}

The success of the above numerical simulation in reproducing an actual physical interference pattern leads us to discuss how our model may be generalized for massive particles, i.e. how one may generate an interference pattern from an ensemble of individual non-interacting massive particles. It is interesting that, for massless photons, there was no mention whatsoever of the Planck constant $h$, except for the fact that the energy $\epsilon$ of a photon is equal to $hc/\lambda$, where $c$ is the speed of light. This result is very promising because we knew about the wave character of electromagnetic radiation well before the discovery of quantum mechanical interference. The new result that quantum mechanics brought in the discussion is that electromagnetic radiation consists of a sequence of individual photons. We have shown above that, for our proposed special type of detector, individual non-interacting photons too manifest wave characteristics without invoking quantum mechanics.

All the above discussion is directly generalizable for massive particles of mass $m$ and velocity ${\rm d}{\bf x}/{\rm d}t$ if we replace the photon wavelength $\lambda$ with the de Broglie wavelength $\lambda_{\rm de\ Broglie}\equiv \hbar/(m|{\rm d}{\bf x}/{\rm d}t|)$, and the phase of the massive particle with
\begin{equation}
\phi \equiv \int \frac{2\pi {\rm d}{\rm l}}{\lambda_{\rm de\ Broglie}} = \int \frac{m}{\hbar}\frac{{\rm d}{\bf x}}{{\rm d}t}\cdot {\rm d}{\bf x}
\ .
\label{phiparticle}
\end{equation} 
Here, the ${\rm d}{\bf x}$ integration is taken along the particle trajectory. The reason we need an integration is because the magnitude of the particle velocity 
$|{\rm d}{\bf x}/{\rm d}t|$ is in general not constant, and thus the de Broglie wavelength may vary along the particle trajectory. On the contrary, a photon's velocity always remains fixed, and the above integration yields $2\pi{\rm l}/\lambda$ (eq.~\ref{phiphoton}). Eqs.~(\ref{Psi}) and (\ref{N}) are directly generalizable as
\begin{equation}
\overline{\Psi}({\bf x};t) \equiv \sum\nolimits_{k=1}^K\frac{1}{\sqrt{N_k}}\sum\nolimits_{j=1}^{N_k} {\rm e}^{i \int_{k} m\frac{{\rm d}{\bf x}_j}{{\rm d}t}\cdot {\rm d}{\bf x}/\hbar}\ ,
\label{Psimassive}
\end{equation}
\begin{equation}
\overline{N}\equiv |\overline{\Psi}|^2({\bf x};t)=\left|\sum\nolimits_{k=1}^K \frac{1}{\sqrt{N_k}}\sum\nolimits_{j=1}^{N_k} {\rm e}^{i \int_k m\frac{{\rm d}{\bf x}_j}{{\rm d}t}\cdot {\rm d}{\bf x}/\hbar}\right|^2\ .
\label{Nmassive}
\end{equation}
Once again, the integrals $\int_k$ are calculated along the $K$ possible different trajectories that particles may follow to reach the spacetime position $({\bf x};t)$, $N_1+\ldots +N_K=N$ is the total number of particles collected by the detector, and in general $\overline{N}\neq N$. From the above discussion, it seems that the de Broglie wavelength may be equally fundamental as Planck's constant $h$. 

We soon realized that, in certain physical systems, massive particles cannot follow classical trajectories with conserved energy $\epsilon$ and at the same time be distributed according to a solution of the Schrodinger equation. Quantum tunneling is one such example. Another example is the hydrogen atom where an electron say at its ground state with enery $\epsilon_1$ can be found at all distances away from the nucleus (the corresponding wave function extends all the way to infinity), although this is impossible for a classical particle with negative energy $\epsilon_1$ that is under the action of the Coulomb potential. Obviously, we are missing something important.

Nelson~(1966) first proposed that microscopic particles may follow frictionless Brownian motion. According to his proposition, particles of mass $m$ interact with a background with diffusion coefficiant $\nu$ and no friction. This stochastic motion introduces randomness in the determination of the phase of a particle along its trajectory. Nelson proposed the following choice of wavefunction,
\begin{equation}
\Psi_{\rm Nelson} = \sqrt{n}\ {\rm e}^{i {\cal S}}
\equiv \sqrt{n}\ {\rm e}^{i\int m{\rm {\bf v}}\cdot {\rm d}{\bf x}/\hbar}\ ,
\label{psiNelson}
\end{equation}
where, ${\rm {\bf v}}$ is the so-called particle current velocity, and $n$ is the density of detected particles at a particular spacetime position. Let us backtrack a little and present to the reader a simplified formulation of Brownian motion similar to the one used by Nelson. Let ${\bf x}(t)$ be a stochastic process that describes the particle position with time $t$ and satisfies
\begin{equation}
{\rm d}{\bf x}(t)={\rm {\bf v}}\ {\rm d}t+{\rm d}{\bf w}(t)\ ,
\label{dx}
\end{equation}
where ${\bf w}$ is a Wiener process, and ${\rm {\bf v}}$ is the mean particle velocity. The ${\rm d}{\bf w}$ are independent Gaussian with mean $\langle{\rm d}{\bf w}_i(t)\rangle=0$, and
$\langle{\rm d}{\bf w}_i(t)\cdot{\rm d}{\bf w}_j(t)\rangle=
2\nu \delta_{ij} {\rm d}t$. In a stochastic process, ${\bf x}(t)$ is not differentiable, thus we need to take proper care of time derivatives. In an average sense, one may integrate the following set of equations,
\begin{eqnarray}
\langle\frac{{\rm d}{\rm {\bf x}}}{{\rm d}t}\rangle & = & {\bf v}
\label{Nelson1}\\
\frac{\partial{\rm {\bf v}}}{\partial t} & = & 
\frac{{\bf F}}{m} - ({\bf v}\cdot \nabla){\bf v}+({\bf u}\cdot \nabla){\bf u}+\nu\Delta{\bf u}\\
\frac{\partial{\rm {\bf u}}}{\partial t} & = & 
-\nu \nabla(\nabla\cdot {\bf v})-\nabla({\bf v}\cdot{\bf u})\ ,
\label{Nelson3}
\end{eqnarray}
and obtain the distributions of ${\rm {\bf v}}$ and $n$ from them. Here, ${\bf F}\equiv -\nabla V$ is the external force acting on the particle, and ${\bf u}\equiv \nu \nabla \ln n$ is the so-called osmotic velocity. The final assumptions are that ${\bf v}$ is a gradient, namely
\begin{equation}
{\bf v}\equiv \frac{\hbar}{m}\nabla{\cal S}\ ,
\label{vNelson}
\end{equation}
and that the diffusion coefficient is equal to
\begin{equation}
\nu=\frac{\hbar}{2m}\ .
\label{nu}
\end{equation}
Under the above assumptions, Nelson was able to show that the expression for $\Psi_{\rm Nelson}$ in eq.~(\ref{psiNelson}) satisfies Schrodinger's equation.

The new element brought into the discussion by our model of non-noninteracting particles is that $\Psi_{\rm Nelson}$ may somehow be related to our function $\overline{\Psi}$. $\overline{\Psi}$ is not carried by each individual particle, but is calculated by a detector at a certain position of the experiment. According to eqs.~(\ref{Psimassive}) and (\ref{dx}),
\begin{equation}
\overline{\Psi} =\sum\nolimits_{k=1}^K\frac{{\rm e}^{i\int_k m{\rm {\bf v}_k}\cdot {\rm d}{\bf x}/\hbar}}{\sqrt{N_k}}\sum\nolimits_{j=1}^{N_k}{\rm e}^{i\int_k m\frac{{\rm d}{\bf w}_j}{{\rm d}t}\cdot {\rm d}{\bf x}/\hbar}\ ,
\end{equation}
where again, the first summation is over all the different discrete trajectories that particles may follow to reach the detector at spacetime position $({\bf x};t)$. According to eq.~(\ref{Nmassive}), $|\Psi|^2$ will be equal to the number $\overline{N}$ of detected particles at a particular detection region of spacetime. If we ignore for the moment the stochastic perturbations ${\rm d}{\bf w}_j$, $\overline{\Psi}$ becomes
\begin{equation}
\overline{\Psi} = \sum\nolimits_{k=1}^K \sqrt{N_k}\ {\rm e}^{i\int_k m{\bf v}_k\cdot {\rm d}{\bf x}/\hbar}\ .
\end{equation}
When we divide by the square root of the volume of space $\delta {\cal V}$ where the detection takes place we obtain
\begin{eqnarray}
\Psi_{\rm Nelson}&\equiv& \frac{\overline{\Psi}}{\sqrt{\delta {\cal V}}}
\equiv
\sqrt{n}\ {\rm e}^{i\int m{\bf v}\cdot {\rm d}{\bf x}/\hbar}
\nonumber\\
&=&\sum\nolimits_{k=1}^K \sqrt{n_k}\ {\rm e}^{i\int_k m{\bf v}_k\cdot {\rm d}{\bf x}/\hbar}
\ ,
\label{Nelsonsimple}
\end{eqnarray}
where $n_k\equiv N_k/\delta {\cal V}$ are the partial densities of particles that reached the detection point following path $k$, and the integrations in the last term are along the directions of the particle current velocities ${\bf v}_k$. As we emphasized over and over again, in general $\sum\nolimits_k n_k \neq n\equiv \overline{N}/\delta {\cal V}$. Notice that ${\bf v}$ as defined in eq.~(\ref{vNelson}) is the velocity of particles in Bohmian mechanics \cite{6}. In the case of multiple particle trajectories, however, our model of non-interacting particles assumes that individual particles {\it do not} flow along ${\bf v}$, but follow their own classical trajectories with average velocities ${\bf v}_k\neq {\bf v}$.

Nelson's derivation of the Schrodinger equation requires the action of a Wiener process ${\bf w}$ with diffusion coefficient $\nu=\hbar/2m$, and the integration of eqs.~(\ref{Nelson1}-\ref{Nelson3}). We here propose that the wavefunction may be constructed as
\[
\Psi_{\rm Nelson} \equiv \sqrt{n}\ {\rm e}^{i\int m{\rm {\bf v}}\cdot {\rm d}{\bf x}/\hbar}\equiv \frac{\overline{\Psi}}{\sqrt{\delta {\cal V}}}
\]
\begin{equation}
=
\sum\nolimits_{k=1}^K \sqrt{n_k}\ {\rm e}^{i\int_k m{\rm {\bf v}_k}\cdot {\rm d}{\bf x}/\hbar}\left(\frac{\sum\nolimits_{j=1}^{N_k}{\rm e}^{i\int_k m\frac{{\rm d}{\bf w}_j}{{\rm d}t}\cdot {\rm d}{\bf x}/\hbar}}{N_k}\right)
\label{Nelsonfinal}
\end{equation}
where the integrations are performed along the independent stochastic trajectories that individual particles follow from their position of origin to the position of their eventual detection after a large number of repetitions of the experiment. We can only get a hint here about how the random phases in the above expressions behave after a large number of repetitions of the experiment. Their average value is equal to
\begin{eqnarray}
 \langle \int \frac{m}{\hbar}\frac{{\rm d}{\bf w}_j}{{\rm d}t}
\cdot {\rm d}{\bf x}  \rangle &\approx& \sum\nolimits  \frac{m}{\hbar\ {\rm d}t}\ \langle{\rm d}{\bf w}_j\cdot{\rm d}{\bf w}_j\rangle \nonumber\\
&=& 
   \sum\nolimits \frac{m\  2\nu}{\hbar}= \sum\nolimits 1 = {\cal M}\ ,
 \end{eqnarray}
where ${\cal M}$ is the number of stochastic kicks that each particle experiences along its trajectory. In other words, particles collected at the same time instance $t$ in eq.~(\ref{Nelsonfinal}), all of them experience the same number ${\cal M}$ of stochastic kicks, and therefore, the term in parentheses in eq.~(\ref{Nelsonfinal}) is expected to yield only an unimportant extra phase from the contribution of these stochastic kicks. This result remains to be shown with an actual numerical experiment that will be performed in a followup publication. 

In a world where all measurements are performed according to our model of a special type of detector, an interference pattern is not something that characterizes each individual particle. Particles have no a-priori knowledge of their distribution in space and time during a large number of repetitions of the experiment. This is in contrast to other non-local particle theories like Bohmian mechanics \cite{6}. Particles carry their own phases,  follow classical trajectories that are acted upon by external forces, and are stochastically perturbed according to eq.~(\ref{dx}) everytime they interact with a background, or every time we decide to make a measurement on them. This is all there is to it. The particles do not know anything a-priori about their distribution $n$. In order to determine their distribution after several repetitions of the experiment, we need to place detectors at all positions in space and time and collect detections over and over again. The role of the detector is to collect the phases of the particles that reach it, and process them according eq.~(\ref{Psi}). Then we can gradually build a spatial density distribution $n$ that is proportional to the square of the amplitude of the function $\overline{\Psi}$.

We cannot emphasize more this result. In our special world, individual particles are not associated with any wave. Wave-like behavior emerges from classical orbits of non-interacting particles perturbed by a large number of stochastic interactions. This is how $n_k$ and ${\bf v}_k$ are determined and from them $n$, ${\bf v}$ and $\Psi_{\rm Nelson}$ in eq.~(\ref{Nelsonsimple}), not the other way around by first solving Schrodinger's equation.

\section{\label{sec:level1}A speculative new interpretation of quantum mechanics}

As we will see in a followup publication, our individual particle model holds great promise in reproducing the quantum wavefunction from local first principles. This leads us to propose a new interpretation of non-relativistic quantum mechanics according to which particles characterized by their mass $m$, velocity ${\rm d}{\rm {\bf x}}/{\rm d}t$, and phase $\phi$ follow classical Brownian trajectories.
When these particles start to move, they receive some kind of `kick' that `sets-off' an `internal clock' that records the evolution of their phase according to eq.~(\ref{phiparticle}). In other words, it is the transfer of energy and momentum to a particle that also triggers its internal phase. An analogy may be made with a soccer ball which, when kicked, travels in space but also oscillates internally. The same thing will happen also every time the particle interacts stochastically with a detector. 

When particles reach a detector, the detector records their phases and processes them according to eq.~(\ref{Psimassive}). Particles exist and behave classically, but manifest themselves only when they interact with a detector, or equivalently their macroscopic classical environment. The detector records a detection when the square of $\overline{\Psi}$ obtained from several repetitions of the experiment reaches a certain detection threshold. In other words, individual particles exist and travel along stochastic classical trajectories, it is only when we decide to detect them that we observe wave-like characteristics. Nature behaves classically, and wave-like behavior manifests itself only in the eyes of the detector.
This interpretation is intriguing
and we plan to continue investigating it.

\paragraph*{\bf Acknowledgements:}
We acknowledge extensive corrections by Pr. George Contopoulos which greatly improved the presentation of our ideas.


\begin{thebibliography}{06}
\bibitem{1} Ball, P., Nature, 12198 (2013)
\bibitem{2} Basi, A. et al., Reviews of Modern Physics, {\bf 85}, 471 (2013)
\bibitem{3} Bell, J., Physics World, {\bf 3}, 33 (1990)
\bibitem{4} Feynman, R. P., Leighton, R. B., Sands, M., `The Feynman Lectures on Physics. Volume III, Addison-Wesley, Reading (1965)
\bibitem{5} Jin, F., Yuan, S., De Raedt, H., Michielsen, K., Miyashita, S., Journal  of the Physical Society of Japan, {\bf 79}, 074401 (2010)
\bibitem{6} Goldstein, S., in Stanford Encyclopedia of Philosophy, Zalt. E. N. (ed.), Stanford Univ. (2017)
\bibitem{7} Harrigan, N. R., Spekkens, R. W., Foundations of Physics, {\bf 40}, 125 (2010)
\bibitem{8} Kolenderski, P. et al., Scientific Reports, {\bf 4}, 4685 (2014)
\bibitem{9} Nelson, E., Physical Review, {\bf 150}, 1079 (1966)
\bibitem{Jin} Jin et al.~(2010) propose to process the phases of individual photon detections as $\overline{\Psi}_N\equiv \gamma \overline{\Psi}_{N-1}+(1-\gamma){\rm e}^{i\phi_N}=(1-\gamma)\sum\nolimits_{j=1}^N \gamma^{N-j} {\rm e}^{i\phi_j}$, where $0<\gamma<1$ is one parameter of their model. This procedure differs from our eq.~(\ref{Psi}). In their model, $0<|\overline{\Psi}|_N<1$, whereas in ours it is of order $\sqrt{N}$, i.e. it grows indefinitely as more and more photons are collected in subsequent realizations of the experiment. In order to trigger their detector, they compare $|\overline{\Psi}|^2_N$ with a random number $r_N$ uniformly distributed betwen 0 and 1. Although both detector models generate wave-like interference patterns, we believe that our approach is more physical. This is why we believe there is potential to use it as a new interpretation of quantum mechanical measurement.
\end{thebibliography}
\end{document}